\documentclass[a4paper,twoside]{article}

\usepackage{epsfig}
\usepackage{subfigure}
\usepackage{calc}
\usepackage{amssymb}
\usepackage{amstext}
\usepackage{amsmath}
\usepackage{amsthm}
\usepackage{multicol}
\usepackage{pslatex}
\usepackage{apalike}
\usepackage{SCITEPRESS}
\usepackage[small]{caption}

\begin{document}

\title{JETracer - A Framework for Java GUI Event Tracing}

\author{\authorname{Arthur-Jozsef Molnar}
\affiliation{Faculty of Mathematics and Computer Science, Babe\c{s}-Bolyai University, Cluj-Napoca, Romania}
\email{arthur@cs.ubbcluj.ro}
}


\abstract{The present paper introduces the open-source Java Event Tracer (JETracer) framework for real-time tracing of GUI events within applications based on the AWT, Swing or SWT graphical toolkits. Our framework provides a common event model for supported toolkits, the possibility of receiving GUI events in real-time, good performance in the case of complex target applications and the possibility of deployment over a network. The present paper provides the rationale for JETracer, presents related research and details its technical implementation. An empirical evaluation where JETracer is used to trace GUI events  within five popular, open-source applications is also presented.}

\onecolumn \maketitle \normalsize \vfill

\section{\uppercase{Introduction}}
\label{sec:Introduction}
The graphical user interface (GUI) is currently the most pervasive paradigm for human-computer interaction. With the continued proliferation of mobile devices, GUI-driven applications remain the norm in today's landscape of pervasive computing. Given their virtual omnipresence and increasing complexity across different platforms, it stands to reason that tools supporting their lifecycle must keep pace. This is especially evident for many applications where GUI-related code takes up as much as 50\% of all application code \cite{Memon01}.

Therefore we believe that software tooling that supports the GUI application development lifecycle will take on an increasing importance in practitioners' toolboxes. Such software can assist with many activities, starting with analysis and design, as well as coding, program comprehension, software visualization and  testing. This is confirmed when studying the evolution of a widely used IDE such as Eclipse \cite{Hou09}, where each new version ships with more advanced features which are aimed at helping professionals create higher quality software faster. Furthermore, the creation of innovative tools is nowadays aided by the prevalence of managed, flexible platforms such as Java and .NET, which enable novel tool approaches via techniques such as reflection, code instrumentation and the use of annotations.

The supportive role of tooling is already well established in the literature. In \cite{Maalej14}, the authors conduct an industry survey covering over 1400 professional developers regarding the strategies, tools and problems encountered by professionals when comprehending software. Among the most significant findings are that developers usually interact with the target application's GUI for finding the starting point of further interaction as well as the use of IDE's in parallel with more specialized tools. Of particular note was the finding that "industry developers do not use dedicated program comprehension tools developed by the research community" \cite{Maalej14}. Another important issue regards open access to state of the art tooling. As we show in the following section, there exist commercial tools that incorporate some of the functionalities of JETracer. However, as they are closed-source and available against significant licensing fees, they have limited impact within the academia.

Our motivation for developing JETracer is the lack of open-source software tools providing multi-platform GUI event tracing. We believe GUI event tracing supports the development of further innovative tools. These may target program comprehension or visualization by creating runtime traces or software testing by providing real-time information about executed code.

The present paper is structured as follows: the next section presents related work, while the third section details JETracer's technical implementation. The fourth section is dedicated to an evaluation using 5 popular open-source applications. The final section presents our conclusions together with plans for future work.

\section{\uppercase{Related Work}}
\label{sec:RelatedWork}
The first important work is the Valgrind\footnote{http://valgrind.org/} multi-platform dynamic binary instrumentation framework. Valgrind loads the target into memory and instruments it \cite{Nethercote07} in a manner that is similar with our approach. Among Valgrind's users we mention the DAIKON invariant detection system \cite{Perkins04} as well as the TaintCheck system \cite{Newsome05}. An approach related to Valgrind is the DTrace\footnote{http://dtrace.org/blogs} tool. Described as an "observability technology" by its authors \cite{Cooper12}, DTrace allows observing what system components are doing during program execution

While the first efforts targeted natively-compiled languages from the C family, the prevalence of instrumentation-friendly and object oriented platforms such as Java and .NET spearheaded the creation of supportive tooling from platform developers and third parties alike. In this regard we mention Oracle's Java Mission Control and Flight Recorder tools \cite{Oracle13} that provide monitoring for Java systems. Another important contribution is Javaassist, a tool which facilitates instrumentation of Java class files, including core classes during JVM class loading \cite{Chiba04}. Its capabilities and ease of use led to its widespread use in dynamic analysis and tracing \cite{Merwe14}. As discussed in more detail within the following sections, JETracer uses Javaassist for instrumenting key classes responsible for firing events within targeted GUI frameworks.

The previously detailed frameworks and tools have facilitated the implementation of novel software used both in research and industry targeting program comprehension, software visualization and testing. A first effort in this direction was the JOVE tool for software visualization \cite{Reiss05}. JOVE uses code instrumentation to capture snapshots of each working thread and create a program trace which can then be displayed using several visualizations \cite{Reiss05}. A different approach is taken within Whyline, which proposes a number of "Why/Why not" type of questions about the target program's textual or graphical output \cite{Ko2010}. Whyline uses bytecode instrumentation to create program traces and record the program's graphical output, with the execution history used for providing answers to the selected questions.

An important area of research where JETracer is expected to contribute targets the visualization of GUI-driven applications. This is of particular interest as an area where research results recently underwent large-scale industrial evaluation with encouraging results \cite{Pekka14}. A representative approach is the GUISurfer tool, which builds a language independent model of the targeted GUI using static analysis \cite{Silva10}.

Another active area of research relevant to JETracer is GUI application testing, an area with notable results both from commercial as well as academic organizations. The first wave of tools enabling automated testing for GUI applications are represented by capture-replay implementations such as Pounder or Marathon, which enable recording a user's interaction with the application \cite{Nedyalkova13}. The recorded actions are then replayed automatically and any change in the response of the target application, such as an uncaught exception or an unexpected window being displayed are interpreted as errors. The main limitation of such tools lays in limitations when identifying graphical widgets, as changes to the target application can easily break test case replayability. More advanced tools integrate scripting engines facilitating quick test suite creation such as Abbot and TestNG \cite{Ruiz07}. However, existing open-source tools are limited to a single GUI toolkit \cite{Nedyalkova13}. Even more so, some of these tools such as Abbot and Pounder are no longer in active development, and using them with the latest version of Java yields runtime errors. 

These projects paved the way for commercial implementations such as MarathonITE\footnote{http://marathontesting.com/}, a fully-featured and commercial implementation of Marathon or the Squish\footnote{http://www.froglogic.com/squish} toolkit. When compared with their open-source alternatives, these applications provide greater flexibility by supporting many GUI toolkits such as AWT/Swing, SWT, Qt, Web as well as mobile platforms. In addition, they provide more precise widget recognition algorithms which helps with test case playback. GUI interactions can be recorded as scripts using non-proprietary languages such as Python or JavaScript, making it easier to modify or update test cases. As part of our research we employed the Squish tool for recording consistently replayable user interaction scenarios which are described in detail within the evaluation section. From the related work surveyed, we found the Squish implementation to be the closest one to JETracer. The Squish tool consists of a server component that is contained within the IDE and a hook component deployed within the AUT \cite{Froglogic15}. The Java implementation of Squish uses a Java agent and employs a similar architecture to our own framework, which is detailed within the following section. 

In contrast to commercial implementations encumbered by restrictive and pricey licensing agreements, we designed JETracer as an open framework to which many tools can be connected without significant programming effort. Furthermore, JETracer facilitates code reuse and a modular approach so that interested parties can add modules to support other GUI toolkits.

The last, but most important body of research addressing the issue of GUI testing has resulted in the GUITAR framework \cite{Nguyen13}, which includes the GUIRipper \cite{Memon13} and MobiGUITAR \cite{Domenico14} components able to reverse engineer desktop and mobile device GUIs, respectively. Once the GUI model is available, valid event sequences can be modelled using an event-flow graph or event-dependency graph \cite{Yuan10,Arlt12}. Information about valid event sequences allows for automated test case generation and execution, which are also provided in GUITAR \cite{Nguyen13,Domenico14}. 

The importance of the GUITAR framework for our research is underlined by its positive evaluation in an industrial context \cite{Pekka13,Pekka14}. While GUITAR and JETracer are not integrated, JETracer's creation was partially inspired by limitations within GUITAR caused by its implementation as a black-box toolset. One of our future avenues of research consists in integrating the JETracer framework into GUITAR and using the event information available to further guide test generation and execution in a white-box process.

\section{\uppercase{The JETracer Framework}}
\label{sec:JETracerFramework}
\begin{figure}[!h]
  \centering
   {\epsfig{file = 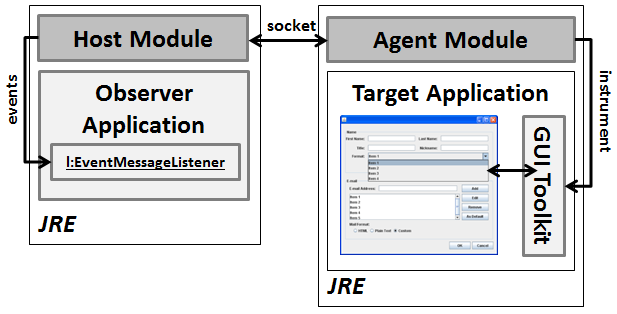, width = 7.5cm}}
  \caption{JETracer deployment architecture}
  \label{fig:example1}
\end{figure}

JETRacer is provided under the Eclipse Public License and is freely available for download from our website \cite{JETracer15}. The implementation was tested under Windows and Ubuntu Linux, using versions 6, 7 and 8 of both Oracle and OpenJDK Java. JETracer consists of two main modules: the \emph{Host Module} and the \emph{Agent Module}. The agent module must be deployed within the target application's classpath. The agent's role is to record the fired events as they occur and transmit them to the host via network socket, while the host manages the network connection and transmits received events to subscribed handlers. JETracer's deployment architecture within a target application is shown in Figure \ref{fig:example1}.

To deploy the framework, the \emph{Agent Module} must be added to the target application's classpath, while the \emph{Host Module} must exist on the classpath of the application interested in receiving event information, which is illustrated in Figure \ref{fig:example1} as the \emph{Observer Application}. Since communication between the modules happens via network socket, the target application and the host module need not be on the same physical machine. 

The framework can be extended to provide event tracing for other GUI toolkits. Interested parties must develop a new library to intercept events within the targeted GUI toolkit and transform them into the JETracer common model. Within this model, each event is represented by an instance of the \emph{EventMessage} class. As it is this common model instance that is transmitted via network to the host, the \emph{Host Module} implementation does not depend on the agent, which allows reusing the host for all agents.

\subsection{The \emph{Host Module}}
The role of the host module is to transparently manage the connection with the deployed agent, to configure the agent and to receive and forward fired events. The code snippet below illustrates how to initialize the host module within an application in order to establish a connection with an agent:

\begin{small}
\begin{verbatim}
 InstrService is = new InstrService();
 InstrConnector connector = is.configure(config);
 connector.connect();
 connector.addEventMessageListener(this);
\end{verbatim}
\end{small}

The kind, type and granularity of recorded events can be filtered using an instance of the \emph{InstrumentConfig} class as detailed within the section dedicated to the agent module. The object is passed as the single parameter of the \emph{configure(config)} method in the code snippet above. In order to receive fired events, at least one \emph{EventMessageListener} must be created and registered with the host, as shown above. The notification mechanism is implemented according to Java best practices, with the \emph{EventMessageListener} interface having just one method:

\begin{small}
\begin{verbatim}
 messageReceived(EventMessageReceivedEvent e);
\end{verbatim}
\end{small}

The received \emph{EventMessageReceivedEvent} instance wraps one \emph{EventMessage} object which describes a single GUI event using the following information:

\begin{itemize}
    \item\emph{Id}: Unique for each event.
    \item\emph{Class}: Class of the originating GUI component (e.g. \emph{javax.swing.JButton})
    \item\emph{Index}: Index in the parent container of the originating GUI component.
    \item\emph{X,Y, Width, Height}: Location and size of the GUI component which fired the event.
	\item\emph{Screenshot}: An image of the target application's active window at the time the event was fired.
	\item\emph{Type}: The type of the fired event (e.g. \emph{java.awt.event.ActionEvent}).
	\item\emph{Timers}: The values for a number of high-precision timers for measurement purposes.
	\item\emph{Listeners}: The list of event handlers registered within the GUI component that fired the event.
\end{itemize}

We believe that the data gathered by JETracer opens the possibility for undertaking a wide variety of analyses. Recording screenshots together with component location and size allows actioned widgets to be identified visually. Likewise, recording each component's index within their parent enables them to be identified programmatically, which can help in creating replayable test cases \cite{McMaster09}. Knowledge about each component's event listeners gathered at runtime has important implications for program comprehension as well as white-box testing by showing how GUI components and the underlying code are linked \cite{Molnar12}. 

\subsection{The \emph{Agent Module}}
The role of the agent module is to record fired events as they happen, gather event information and transmit it to the host. The existing agents do most of the work during class loading, when they instrument several classes from the Java platform using Javaassist. The actual methods that are instrumented are toolkit-specific, but a common approach is employed. In the first phase, we studied the publicly available source code of the Java platform and identified the methods responsible for firing events. Code which calls our event-recording and transmission code was inserted at the beginning and end of each such method. The code that is instrumented belongs to the Java platform itself, which enables deploying JETracer for all applications that use or extend standard GUI controls.

As GUI toolkits generate a high number of events, excluding uninteresting events from tracing becomes important in order to avoid impacting target application performance. This is achieved in JETracer by applying the following filters:

\emph{Event granularity}: Provides the possibility of recording either all GUI events or only those that have application-defined handlers triggered by the event. This filter allows tracing only those events that cause application code to run.

\emph{Event filter}: Used to ignore certain event types. For example, the AWT/Swing agent records both low and high level events. Therefore, a key press is recorded as three consecutive events: a \emph{KeyEvent.KEY\_PRESSED}, followed by a \emph{KeyEvent.KEY\_RELEASED} and a \emph{KeyEvent.KEY\_TYPED}. If undesirable, this can be avoided by filtering the unwanted events. In our empirical evaluation, we observed that ignoring certain classes of events such as mouse movement and repaint events clear the recorded trace of many superfluous entries and increase target application performance.

Due to differences between GUI toolkits, the AWT/Swing and SWT agents have distinct implementations. As such, our website \cite{JETracer15} holds two agent versions: one that works for AWT/Swing applications and one that works for SWT. A common part exists for maintaining the communication with the host and providing support for code instrumentation, which we hopefully will enable interested contributors to extend JETracer for other GUI toolkits. The sections below detail the particularities of each agent implementation. The complete list of events traceable for each of the toolkits is available on the JETracer project website \cite{JETracer15}.

\subsubsection{The \emph{AWT/Swing Agent}}
Due to the interplay between AWT and Swing we were able to develop one agent module that is capable of tracing events fired within both toolkits. AWT and Swing are written in pure Java and are included with the default platform distribution, so in order to record events we instrumented a number of classes within the \emph{java.awt.*} and \emph{javax.swing.*} packages. This proved to be a laborious undertaking due to the way event dispatch works in these toolkits, as many components have their own code for firing events as well as maintaining lists of registered event handlers. This also required laborious testing to ensure that all event types are recorded correctly.

\subsubsection{The \emph{SWT Agent}}
In contrast to AWT and Swing, the SWT toolkit is available within a separate library that provides a bridge between Java and the underlying system's native windowing toolkit. As such, there exist different SWT libraries for each platform as well as architecture. At the time of writing, JETracer was tested with versions 4.0 - 4.4 of SWT under both Windows and Ubuntu Linux operating systems.

In order to trace events, we have instrumented the \emph{org.eclipse.swt.widgets.EventTable} class, which handles firing events within the toolkit \cite{Northover04}. 

\section{\uppercase{Empirical Evaluation}}
\label{sec:EmpiricalEvaluation}
The present section details our evaluation of the JETracer framework. Our goal is to evaluate the feasibility of deploying JETracer within complex applications and to examine the performance impact its various settings have on the target application. GUI applications are event-driven systems that employ callbacks into user code to implement most functionalities. While GUI toolkits provide a set of graphical widgets together with associated events, applications typically use only a subset of them. Furthermore, applications are free to (un)register event handlers and to update them during program execution. This variability is one of the main issues making GUI application comprehension, visualization and testing difficult. As JETracer captures events fired within the application under test, its performance is heavily influenced by factors outside our control. These include the number and types of events that are fired within the application, the number of registered event handlers as well as the network performance between agent and host components. In order to limit external threats to the validity of our evaluation, both modules were hosted on the same physical machine, a modern quad-core desktop computer running Oracle Java 7 and the Windows operating system. We found our results repeatable using different versions of Java on both Windows and Ubuntu Linux. 

\subsection{Target Applications}
Our selection of target applications was guided by a number of criteria. First, we wanted applications that will enable covering most, if not all GUI controls and events present within AWT/Swing and SWT. Second of all, we searched for complex, popular open-source applications that are in active development. Last but not least, we limited our search to applications that were easy to set up and which enabled the creation of consistently replayable interactions.

The selected applications are the aTunes media player, the Azureus torrent client, the FreeMind mind mapping software, the jEdit text editor and the TuxGuitar tablature editor. We used recent versions for each application except Azureus, where due to the inclusion of proprietary code in recent versions an older version was selected. aTunes, FreeMind and jEdit employ AWT and Swing, while Azureus and TuxGuitar use the SWT toolkit. These applications have complex user interfaces that include several windows and many control types, some of which custom created. Several of them have already been used as target applications in evaluating research results. Previous versions of both FreeMind and jEdit were employed in research targeting GUI testing \cite{Yuan10,Arlt12}, while TuxGuitar was used in researching new approaches in teaching software testing at graduate level \cite{Krutz14}.

\subsection{The Evaluation Process}
The evaluation process consisted of first recording and then replaying a user interaction scenario for each of the applications using different settings for JETracer. These scenarios were created to replicate the actions of a live user during an imagined session of using each application and they cover most control types within each target application. Table \ref{tab:ScenarioEvents} illustrates the total number of events, as well as the number of handled events that were generated when running the scenarios. Differences between the number of generated events are explained by the fact that the interaction scenarios were created to be of approximately equal length from a user's perspective; the actual number of fired events is specific to each application.

\begin{table}[h]
\caption{Number of events recorded during the scenario runs}\label{tab:ScenarioEvents} \centering
\begin{tabular}{|l|c|c|}
  \hline
  Application & All Events & Handled Events\\
  \hline
  aTunes 3.1.2 & 155,502 & 1,724 \\
  \hline
  Azureus 2.0.4.0 & 11,149 & 230 \\
  \hline
  FreeMind 1.0.1 & 356,762 & 5,308 \\
  \hline
  jEdit 4.5.2 & 38,708 & 1,940 \\
  \hline
  TuxGuitar 1.0.2 & 13,696 & 1,802 \\
  \hline
  TOTAL & 575,817 & 11,004 \\
  \hline
\end{tabular}
\end{table}

An important issue that affects reliable replay of user interaction sequences is flakiness, or unexpected variations in the generated event sequence due to small changes in application or system state \cite{Memon13b}. For instance, the location of the mouse cursor when starting the target application is important for applications that fire mouse-related events. In order to control flakiness, user scenarios were created to leave the application in the same state in which it was when started. Furthermore, we employed the commercially-available Squish capture and replay solution for recording and replaying scenarios. Each application run resulted in information regarding the event trace captured by JETracer as well as per event overhead data. We compared this event trace with the scripted interaction scenarios in order to ensure that our framework captures all generated events in the correct order. All the artefacts required for replicating our experiments as well as our results in raw form are available on our website \cite{JETracer15}. 

\subsection{Performance Benchmark}
The purpose of this section is to present our initial data concerning the overhead incurred when using JETracer with  various settings. The most important factors affecting performance are the number of traced events and the overhead that is incurred for each event. Our implementation targets achieving constant overhead in order to ensure predictable target application performance.

Each usage scenario was repeated a number of four times in order to assess the impact of those two settings that we observed to impact performance: event granularity and screenshot recording. As GUI toolkits generate events on each mouse move, keystroke and component repaint, tracing all events provides a worst-case baseline for event throughput. During our preliminary testing we found capturing screenshots to be orders of magnitude slower than recording other event parameters, so we also explored its impact on the performance of our selected applications during event tracing.

As such, the four scenarios consist of tracing all GUI events versus those having handlers installed by the application, each time with and without recording screenshots. Overhead was recorded via high-precision timers and only includes the time required for recording event parameters and sending the message to the host via network socket. In order to account for variability within the environment, we eliminated outlier values from our data analysis.

Table \ref{tab:OverheadNOSCR} provides information regarding average per event overhead obtained with screenshot recording turned off. Our data shows that per-event overhead remains consistent at around 0.1ms within all applications, with a slightly higher value when tracing handled events. These higher values are explained by the additional information that is gathered for these events, as several reflection calls are required to record event handler information. Furthermore, standard deviation was in most cases below 0.2ms, showing good performance consistency.

From a subjective perspective, applications instrumented to trace all events did not present any observable slowdown. Due to the fact that FreeMind consistently generated the highest number of events, we will use it for more detailed analysis. Our interaction scenario is around 6 minutes long when replayed by a user. The incurred overhead without screenshot recording and tracing all events was 16.5 seconds. However, as FreeMind fires many GUI events while initializing the application, most of the overhead resulted in slower application startup, followed by consistent application performance. The behaviour of the other applications was consistent with this observation. When only handled events are traced, even application startup speed is undistinguishable from an uninstrumented start.

\begin{table}[h]
\caption{Average overhead per event without screenshot recording (in milliseconds).}\label{tab:OverheadNOSCR} \centering
\begin{tabular}{|l|c|c|}
  \hline
  Event granularity & All & Handled \\
  \hline
  aTunes 3.1.2 & 0.09 & 0.21 \\
  \hline
  Azureus 2.0.4.0 & 0.13 & 0.31 \\
  \hline
  FreeMind 1.0.1 & 0.09 & 0.18 \\
  \hline
  jEdit 4.5.2 & 0.11 & 0.18 \\
  \hline
  TuxGuitar 1.0.2 & 0.13 & 0.22 \\
  \hline
\end{tabular}
\end{table}

The more interesting situation is once screenshot recording is turned on. This has a noticeable impact on JETracer's performance due to JNI interfacing required by the virtual machine to access OS resources. As screenshot recording overhead is dependant on the size of the application window, the main windows of all applications were resized to similar dimensions taking care not to affect the quality of user interaction. Table \ref{tab:OverheadSCR} details the results obtained with screenshot recording enabled.

\begin{table}[h]
\caption{Average overhead per event with screenshot recording (in milliseconds).}\label{tab:OverheadSCR} \centering
\begin{tabular}{|l|c|c|}
  \hline
  Event granularity & All & Handled \\
  \hline
  aTunes 3.1.2 & 1.78 & 23.21 \\
  \hline
  Azureus 2.0.4.0 & 31.77 & 34.07 \\
  \hline
  FreeMind 1.0.1 & 2.07 & 27.95 \\
  \hline
  jEdit 4.5.2 & 10.10 & 31.09 \\
  \hline
  TuxGuitar 1.0.2 & 28.02 & 31.96 \\
  \hline
\end{tabular}
\end{table}

Our first observation regards the variability in the observed overhead when tracing all events. This is due to the different initialization sequences of the applications. We found that both aTunes and FreeMind do a lot of work on startup, and since at this point their GUI is not yet visible and so screenshots are not recorded, this lowers the reported average value. These events must still be traced however, as they are no different to events fired once the GUI is displayed. The situation is much more balanced once only handled events are traced, in which case we observe that all applications present similar overhead, between 20 and 30ms per event.

Subjectively, turning screenshot recording on resulted in moderate performance degradation when tracing handled events, as the applications became less responsive to user input. In the case of FreeMind, the overhead added another 55 seconds to our 6 minute interaction scenario, but the application remained usable. As expected, the worst degradation of performance was observed when screenshots were recorded for all events fired. This made all 5 applications unresponsive to user input over periods of several seconds due to the large number of recorded screenshots. To keep our  scripts replayable without error, they had to be adjusted by inserting additional wait commands between steps. In this worst case, added overhead was 6.5 minutes for FreeMind and over 3 minutes for both jEdit and TuxGuitar. This performance hit can be alleviated by further filtering the events to be traced. However, a complete evaluation of this is target application-specific and out of our scope.

\begin{figure}[!h]
  \centering
   {\epsfig{file = 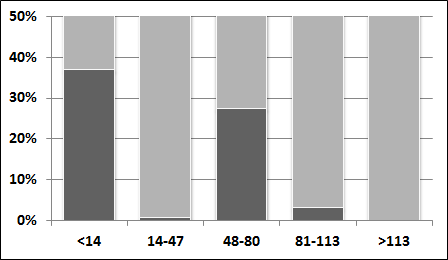, width = 7.5cm}}
  \caption{Distribution of incurred overhead (milliseconds) when tracing handled FreeMind events}
  \label{fig:FreeMindActScr}
\end{figure}

An important aspect regarding target application responsiveness is the consistency of the incurred overhead. As most GUI toolkits are single-thread, inconsistent overhead leads to perceived application slowdown while the GUI is unresponsive. We examined this issue within our selected application, and will report the results for FreeMind, as we found it to be representative for the rest of the applications. First of all, with screenshot recording disabled all events were traced under 1ms, which did not affect application performance. As such, we investigated the issue of consistency once screenshot recording was enabled. Figure \ref{fig:FreeMindActScr} illustrates overhead distribution when tracing handled events. Each column represents one standard deviation from the recorded average of 27.95ms. Overhead clumps into two columns: the leftmost column illustrates events for which screenshots could not be captured as the GUI was not yet visible, while most other events were very close to the mean. 

One of our goals when evaluating JETracer was to compare its performance against other, similar toolkits. However, during our tool survey we were not able to identify similar open-source applications that would enable an objective comparison. Existing applications that incorporate similar functionalities, such as Squish are closed-source so a comparative evaluation was not possible.

\section{\uppercase{Conclusions and Future Work}}
\label{sec:ConclusionsFutureWork}
We envision JETracer as a useful tool for both academia and the industry. We aim for our future work to reflect this by extending JETracer to cover other toolkits such as JavaFX as well as Java-based mobile platforms such as Android. Second of all, we plan to incorporate knowledge gained within our initial evaluation in order to further reduce the framework's performance impact on target applications. We plan to reduce the screenshot capture overhead as well as to examine possible benefits of implementing asynchronous event transmission between agent and host.

Our plans are to build on the foundation established by JETracer. We aim to develop innovative applications for program comprehension as well as software testing by using JETracer to provide more information about the application under test. We believe that by integrating our framework with already established academic tooling such as the GUITAR framework will enable the creation of new testing methodologies. Furthermore, we aim to contribute to the field of program comprehension by developing software tooling capable of using event traces obtained via JETracer. Integration with existing tools such as EclEmma will allow for the creation of new tools to shift the paradigm from code coverage to event and event-interaction coverage \cite{Yuan11} in the area of GUI-driven applications.

\vfill
\bibliographystyle{apalike}
{\small
\bibliography{JETracer}}

\begin{thebibliography}{}

\bibitem[Aho et~al., 2013]{Pekka13}
Aho, P., Suarez, M., Kanstren, T., and Memon, A. (2013).
\newblock Industrial adoption of automatically extracted {GUI} models for
  testing.
\newblock In {\em Proceedings of the 3rd International Workshop on Experiences
  and Empirical Studies in Software Modelling}. Springer Inc.

\bibitem[Aho et~al., 2014]{Pekka14}
Aho, P., Suarez, M., Kanstren, T., and Memon, A. (2014).
\newblock Murphy tools: Utilizing extracted gui models for industrial software
  testing.
\newblock In {\em The Proceedings of the Testing: Academic \& Industrial
  Conference (TAIC-PART)}. IEEE Computer Society.

\bibitem[Amalfitano et~al., 2014]{Domenico14}
Amalfitano, D., Fasolino, A.~R., Tramontana, P., Ta, B.~D., and Memon, A.~M.
  (2014).
\newblock Mobiguitar -- a tool for automated model-based testing of mobile
  apps.
\newblock {\em IEEE Software}.

\bibitem[Arlt et~al., 2012]{Arlt12}
Arlt, S., Banerjee, I., Bertolini, C., Memon, A.~M., and Schaf, M. (2012).
\newblock Grey-box gui testing: Efficient generation of event sequences.
\newblock {\em Computing Research Repository}, abs/1205.4928.

\bibitem[Chiba, 2004]{Chiba04}
Chiba, S. (2004).
\newblock Javassist: Java bytecode engineering made simple.
\newblock {\em Java Developer Journal}.

\bibitem[Cooper, 2012]{Cooper12}
Cooper, G. (2012).
\newblock Dtrace: Dynamic tracing in {Oracle} {Solaris}, {Mac OS X}, and {Free
  BSD}.
\newblock {\em SIGSOFT Softw. Eng. Notes}, 37(1):34--34.

\bibitem[Froglogic, 2015]{Froglogic15}
Froglogic, G. (2015).
\newblock http://doc.froglogic.com/squish/.

\bibitem[Hou and Wang, 2009]{Hou09}
Hou, D. and Wang, Y. (2009).
\newblock An empirical analysis of the evolution of user-visible features in an
  integrated development environment.
\newblock In {\em Proceedings of the 2009 Conference of the Center for Advanced
  Studies on Collaborative Research}, CASCON '09, pages 122--135, New York, NY,
  USA. ACM.

\bibitem[JETracer, 2015]{JETracer15}
JETracer (2015).
\newblock https://bitbucket.org/arthur486/jetracer.

\bibitem[Ko and Myers, 2010]{Ko2010}
Ko, A.~J. and Myers, B.~A. (2010).
\newblock Extracting and answering why and why not questions about java program
  output.
\newblock {\em ACM Trans. Softw. Eng. Methodol.}, 20(2):4:1--4:36.

\bibitem[Krutz et~al., 2014]{Krutz14}
Krutz, D.~E., Malachowsky, S.~A., and Reichlmayr, T. (2014).
\newblock Using a real world project in a software testing course.
\newblock In {\em Proceedings of the 45th ACM Technical Symposium on Computer
  Science Education}, SIGCSE '14, pages 49--54, New York, NY, USA. ACM.

\bibitem[Maalej et~al., 2014]{Maalej14}
Maalej, W., Tiarks, R., Roehm, T., and Koschke, R. (2014).
\newblock On the comprehension of program comprehension.
\newblock {\em ACM Trans. Softw. Eng. Methodol.}, 23(4):31:1--31:37.

\bibitem[McMaster and Memon, 2009]{McMaster09}
McMaster, S. and Memon, A.~M. (2009).
\newblock An extensible heuristic-based framework for gui test case
  maintenance.
\newblock In {\em Proceedings of the IEEE International Conference on Software
  Testing, Verification, and Validation Workshops}, pages 251--254, Washington,
  DC, USA. IEEE Computer Society.

\bibitem[Memon et~al., 2013]{Memon13}
Memon, A., Banerjee, I., Nguyen, B., and Robbins, B. (2013).
\newblock The first decade of gui ripping: Extensions, applications, and
  broader impacts.
\newblock In {\em Proceedings of the 20th Working Conference on Reverse
  Engineering (WCRE)}. IEEE Press.

\bibitem[Memon, 2001]{Memon01}
Memon, A.~M. (2001).
\newblock {\em A comprehensive framework for testing graphical user
  interfaces}.
\newblock PhD thesis.

\bibitem[Memon and Cohen, 2013]{Memon13b}
Memon, A.~M. and Cohen, M.~B. (2013).
\newblock Automated testing of gui applications: models, tools, and controlling
  flakiness.
\newblock In {\em Proceedings of the 2013 International Conference on Software
  Engineering}, ICSE '13, pages 1479--1480, Piscataway, NJ, USA. IEEE Press.

\bibitem[Molnar, 2012]{Molnar12}
Molnar, A. (2012).
\newblock {jSET} - {Java Software Evolution Tracker}.
\newblock In {\em KEPT-2011 Selected Papers}. Presa Universitara Clujeana, ISSN
  2067-1180.

\bibitem[Nedyalkova and Bernardino, 2013]{Nedyalkova13}
Nedyalkova, S. and Bernardino, J. (2013).
\newblock Open source capture and replay tools comparison.
\newblock In {\em Proceedings of the International C* Conference on Computer
  Science and Software Engineering}, C3S2E '13, pages 117--119, New York, NY,
  USA. ACM.

\bibitem[Nethercote and Seward, 2007]{Nethercote07}
Nethercote, N. and Seward, J. (2007).
\newblock Valgrind: A framework for heavyweight dynamic binary instrumentation.
\newblock In {\em Proceedings of the 2007 ACM SIGPLAN Conference on Programming
  Language Design and Implementation}, PLDI '07, pages 89--100, New York, NY,
  USA. ACM.

\bibitem[Newsome, 2005]{Newsome05}
Newsome, J. (2005).
\newblock Dynamic taint analysis for automatic detection, analysis, and
  signature generation of exploits on commodity software.
\newblock In {\em Internet Society}.

\bibitem[Nguyen et~al., 2013]{Nguyen13}
Nguyen, B.~N., Robbins, B., Banerjee, I., and Memon, A. (2013).
\newblock Guitar: an innovative tool for automated testing of gui-driven
  software.
\newblock {\em Automated Software Engineering}, pages 1--41.

\bibitem[Northover and Wilson, 2004]{Northover04}
Northover, S. and Wilson, M. (2004).
\newblock {\em {SWT: The Standard Widget Toolkit, Volume 1}}.
\newblock Addison-Wesley Professional, first edition.

\bibitem[Oracle, 2013]{Oracle13}
Oracle, C. (2013).
\newblock Advanced {Java Diagnostics and Monitoring Without Performance
  Overhead}.
\newblock Technical report.

\bibitem[Perkins and Ernst, 2004]{Perkins04}
Perkins, J.~H. and Ernst, M.~D. (2004).
\newblock Efficient incremental algorithms for dynamic detection of likely
  invariants.
\newblock {\em SIGSOFT Softw. Eng. Notes}, 29(6):23--32.

\bibitem[Reiss and Renieris, 2005]{Reiss05}
Reiss, S.~P. and Renieris, M. (2005).
\newblock Jove: Java as it happens.
\newblock In {\em Proceedings of the 2005 ACM Symposium on Software
  Visualization}, SoftVis '05, pages 115--124, New York, NY, USA. ACM.

\bibitem[Ruiz and Price, 2007]{Ruiz07}
Ruiz, A. and Price, Y.~W. (2007).
\newblock Test-driven gui development with testng and abbot.
\newblock {\em IEEE Softw.}, 24(3):51--57.

\bibitem[Silva et~al., 2010]{Silva10}
Silva, J.~a.~C., Silva, C., Gon\c{c}alo, R.~D., Saraiva, J.~a., and Campos,
  J.~C. (2010).
\newblock {The GUISurfer Tool: Towards a Language Independent Approach to
  Reverse Engineering GUI Code}.
\newblock In {\em Proceedings of EICS 2010}, pages 181--186, New York, NY, USA.
  ACM.

\bibitem[van~der Merwe et~al., 2014]{Merwe14}
van~der Merwe, H., van~der Merwe, B., and Visser, W. (2014).
\newblock Execution and property specifications for jpf-android.
\newblock {\em SIGSOFT Softw. Eng. Notes}, 39(1):1--5.

\bibitem[Yuan et~al., 2011]{Yuan11}
Yuan, X., Cohen, M.~B., and Memon, A.~M. (2011).
\newblock Gui interaction testing: Incorporating event context.
\newblock {\em IEEE Transactions on Software Engineering}, 37(4):559--574.

\bibitem[Yuan and Memon, 2010]{Yuan10}
Yuan, X. and Memon, A.~M. (2010).
\newblock Generating event sequence-based test cases using gui run-time state
  feedback.
\newblock {\em IEEE Transactions on Software Engineering}, 36(1).

\end{thebibliography}

\vfill
\end{document}